\documentclass[journal]{IEEEtran}
\usepackage{cite}
\usepackage{amsmath,amssymb,amsfonts}
\usepackage{algorithmic}
\usepackage{graphicx}
\usepackage{textcomp}
\usepackage{xcolor}
\usepackage[hyphens]{url}
\usepackage{multirow}
\usepackage{booktabs}
\usepackage{tabularx}
\usepackage{multicol}
\usepackage{makecell}
\usepackage[caption=false,font=normalsize,labelfont=sf,textfont=sf]{subfig}
\usepackage{tikz}
\usepackage{enumitem}
\usepackage[normalem]{ulem}
\usepackage[hidelinks]{hyperref}
\usepackage{multirow}

\title{Phantom: Temporal Caching for Diffusion Models}
\author{Yuzhe Fu, Cong Guo, Zishan Shao, Bowen Duan, Chiyue Wei, Haoxuan Shan,
Ting Jiang, Hancheng Ye, Jonathan Ku, Changchun Zhou, Hai ``Helen'' Li, and Yiran Chen%
\thanks{This work was supported in part by the National Science Foundation under Grant 2625580 and Grant 2616525. (\textit{Corresponding author: Cong Guo.})}%
\thanks{\raggedright Yuzhe Fu is with the Department of Computer Science, Duke University, Durham, NC 27708 USA (e-mail: yuzhe.fu@duke.edu).}%
\thanks{\raggedright Cong Guo, Zishan Shao, Bowen Duan, Chiyue Wei, Haoxuan Shan, Ting Jiang, Hancheng Ye, Jonathan Ku, Changchun Zhou, Hai ``Helen'' Li, and Yiran Chen are with the Department of Electrical and Computer Engineering, Duke University, Durham, NC 27708 USA (e-mail: cong.guo@duke.edu; hai.li@duke.edu; \mbox{yiran.chen@duke.edu}).}}

\newcommand{\T}{\emph{Phantom}} %

\definecolor{mycolorblue}{RGB}{58,142,237}
\definecolor{mycolorred}{RGB}{187,75,94}
\definecolor{mycolorgreen}{RGB}{166,213,95}
\definecolor{mycoloryellow}{RGB}{242,210,98}
\definecolor{mycolorpink}{RGB}{242,96,119}
\definecolor{mycolorpurple}{RGB}{149,90,189}
\definecolor{mycolororange}{RGB}{255,199,7}

\newcommand\circlednumberblue[1]{%
  \begin{tikzpicture}[baseline=(char.base)]
    \node[shape=circle,,fill=mycolorblue,inner sep=1pt] (char) {\textcolor{white}{\scriptsize\sffamily\bfseries#1}};
\end{tikzpicture}}
\newcommand\circlednumberred[1]{%
  \begin{tikzpicture}[baseline=(char.base)]
    \node[shape=circle,,fill=mycolorred,inner sep=1pt] (char) {\textcolor{white}{\scriptsize\sffamily\bfseries#1}};
\end{tikzpicture}}

\newcommand{\retagNoLink}[1]{\hyperref[#1]{\textcolor{brown}{\textbf{#1}}}} %
\newcommand{\retag}[1]{\hyperref[#1]{\textcolor{brown}{\underline{\textbf{#1}}}}}

\begin{document}
\maketitle

\begin{abstract}

Diffusion models have become the dominant paradigm for high-quality image generation, but their iterative inference across tens of timesteps incurs substantial computational overhead. Prior accelerators exploit temporal similarity in diffusion models via differential quantization or output sparsity, yet suffer from increased memory footprint, design complexity, and image degradation.
In this work, we present \textbf{{\T}}, an algorithm--architecture co-designed accelerator built on \emph{temporal caching}. We view temporal similarity as \emph{differential sparsity}: between adjacent timesteps, only a small fraction of tokens (outliers) change significantly. \T~recomputes only these outlier tokens and reuses the rest from a temporal cache.
\T~runs on standard systolic arrays with lightweight additions: a \T~Engine at the global buffer that detects outliers and updates the cache as activations stream through, and a condensed pipeline that maps the resulting sparse updates onto dense systolic computation for FC, convolution, and attention layers.
Evaluations on modern diffusion models show that \T~achieves an average 4.0$\times$ speedup and 5.5$\times$ energy saving over GPUs and prior accelerators while preserving generation quality.

\end{abstract}


\section{Introduction}

\begin{figure*}[t]
    \centering
    \includegraphics[width=0.95\linewidth]{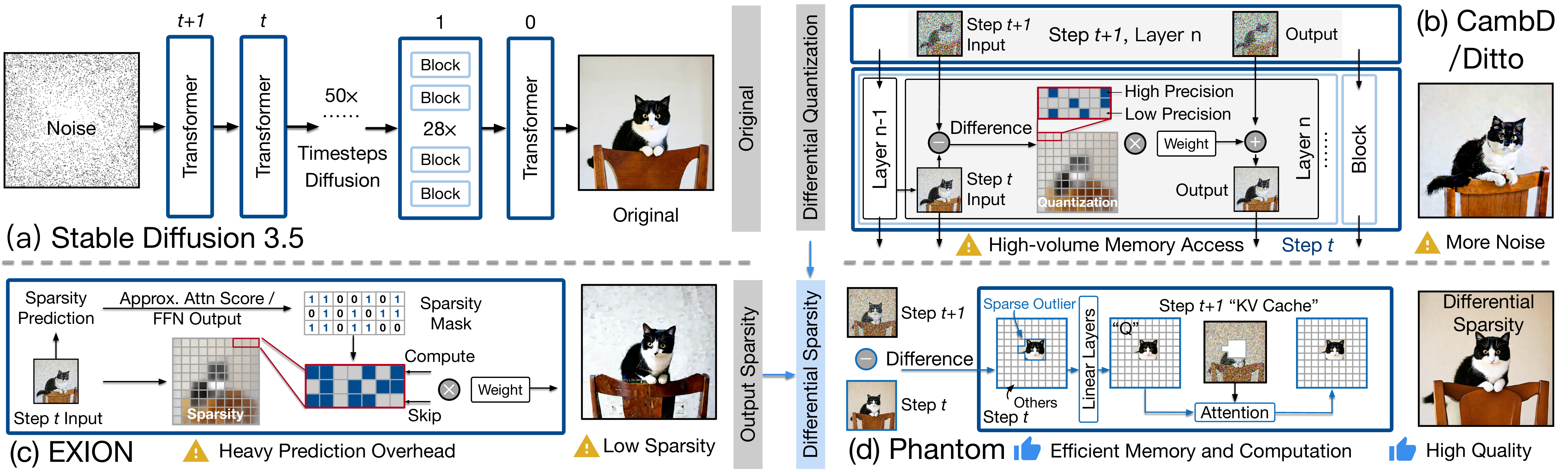}
    \caption{Overview of accelerator architectures for diffusion acceleration. 
    (a) Overall network architecture of Stable Diffusion 3.5 (SD35)~\cite{esser2024scaling}. 
    (b) Differential quantization~\cite{kong2024cambricon,kim2025ditto} and (c) output-sparsity architecture~\cite{heo2025exion} used in prior diffusion accelerators. 
    (d) Our proposed differential sparsity paradigm for diffusion acceleration. 
    Images are generated by SD35 based on MSCOCO dataset~\cite{lin2014microsoft}.}
    \label{fig:intro}
\end{figure*}

Diffusion models have recently emerged as the dominant paradigm for high-quality image generation and power many modern generative systems such as Stable Diffusion~\cite{rombach2022high,esser2024scaling} and DALL-E~\cite{ ramesh2022hierarchical}. By iteratively refining a latent representation from noise to signal, they achieve photorealistic quality with strong controllability, but at substantial computational cost~\cite{navardi2026generative}. As shown in Fig.~\ref{fig:intro}(a), modern diffusion pipelines, like Stable Diffusion 3.5~\cite{esser2024scaling} (SD35), rely on large Transformer-based backbones that must be executed repeatedly across tens of timesteps, making inference compute-intensive~\cite{adaptivediffusion24ye} and challenging for efficient hardware deployment.

Since each timestep performs only a small refinement over the previous state, activations between neighboring steps tend to be highly similar. This observation has inspired both software and hardware optimizations that exploit temporal similarity. One representative approach is \emph{differential quantization} (DQ)~\cite{kong2024cambricon,tang2025diff,kim2025ditto}, illustrated in Fig.~\ref{fig:intro}(b). Instead of computing each timestep independently, these methods compute the difference between consecutive steps and quantize the difference values using mixed precision. Because most differences are small in magnitude, they can be represented using low precision, while large outliers are processed using higher precision. Cambricon-D~\cite{kong2024cambricon} first realized this with INT4/FP16 arithmetic, while Ditto~\cite{kim2025ditto} combines DQ with W8A8 quantization to exploit more low-bit (INT4/INT8) and zero-valued temporal differences.

Differential quantization, however, is fundamentally constrained on two fronts. On the algorithm side, quantization injects numerical noise into a process designed to progressively remove it~\cite{he2023ptqd, li2023qdiffusion}, and since these schemes are training-free and act on dynamic activations, the resulting artifacts become visible on advanced backbones, as shown in Fig.~\ref{fig:intro}(b). 
On the architecture side, the differential representation must be restored to full precision at every non-linear layer, inflating memory traffic. Meanwhile, mixed-precision execution requires both INT and FP units in the same datapath, and this heterogeneity, together with irregular outlier handling, complicates the hardware and limits scalability across evolving models.

Another class of work exploits temporal similarity as sparsity. 
{To limit quality loss, many algorithm works~\cite{wang2024ATEDM, zou2024accelerating, pan2026specdiff} use attention-derived scores to select important tokens for recomputation. 
However, accessing these scores can conflict with fused attention kernels~\cite{dao2024flashattention}, limiting practical speedup: ToCa~\cite{zou2024accelerating} achieves only 0.98$\times$ the FlashAttention baseline on DiT-256.
Some attention-score-free alternatives~\cite{zou2026rethinking,liu2025taylorseers} avoid this dependency but introduce other trade-offs. For example, random token selection~\cite{zou2026rethinking} remains compatible with FlashAttention, but it misses highly changed tokens and yields unstable generation quality (35\% higher FID on DiT~\cite{shao2026soda}).
Existing accelerators do not resolve this tension either: the cost of exposing sparsity still offsets much of the speedup gained from exploiting it.
EXION~\cite{heo2025exion} uses approximate attention score to predict sparsity and reuses previous FFN outputs to skip computation (Fig.~\ref{fig:intro}(c)). However, it requires dedicated prediction hardware and triple-buffered weight storage, with 22\% area overhead in our evaluation. Preserving generation quality on SD35 also requires conservative sparsity settings, limiting exploited sparsity to $\sim$32\%. 
Therefore, effective acceleration requires both a low-cost sparsity detection and an execution path that exploits it with modest hardware and quality costs.}

{To meet both requirements, our key insight is \textbf{\emph{differential sparsity}}:
temporal differences in block inputs serve as a hardware-friendly signal for identifying sparsity. 
As illustrated in Fig.~\ref{fig:intro}(d), these differences are sparse: only a few tokens change significantly across timesteps, so only these tokens require recomputation while the rest can reuse their previous outputs.
Because this signal depends only on block inputs, it can be detected before each block executes with a simple per-token reduction, without accessing attention scores.
It also exposes more reusable computation: on SD35, differential sparsity yields 63\% average sparsity while preserving image quality, approximately twice that exploited by EXION in our evaluation.

Building on this insight, we propose \textbf{\T}, a diffusion accelerator that turns differential sparsity into speedup on standard systolic arrays. 
Rather than relying on heterogeneous mixed-precision datapaths like differential-quantization accelerators~\cite{kong2024cambricon,tang2025diff,kim2025ditto} or a heavy prediction array~\cite{heo2025exion}, \T~detects outlier tokens on the fly and reuses the remaining activations from a \textbf{\emph{temporal cache}}, realized as a partition of the existing global buffer rather than a dedicated structure.
As a result, \T~preserves both the original systolic dataflow and full-precision arithmetic, introducing no quantization noise and sustaining generation quality even on the most complex backbones.}

However, realizing temporal cache on hardware nevertheless poses its own challenges. Unlike the sequential token accumulation of an LLM KV-cache~\cite{cheng2025ecco}, diffusion activations evolve spatially: as the latent is refined, outliers propagate through transformer blocks and the sparsity pattern shifts from layer to layer. Outliers must therefore be re-identified before every block rather than following a fixed pattern, and the corresponding cache entries updated at irregular locations, neither of which a conventional systolic array can support efficiently. 

{\T~addresses this with a lightweight \T~\emph{Engine} placed alongside the systolic arrays, where it detects outliers and updates the temporal cache as activations stream past.} Built on the same datapath, a condensed pipeline compacts the resulting row-sparse activations across FC, convolutional, and attention layers into dense operands, so that the exposed sparsity translates into speedup on an unmodified systolic dataflow.

Experimental results show that \T~achieves an average \textbf{4.0$\times$} speedup and \textbf{5.5$\times$} energy savings across GPU and accelerator baselines, while maintaining near-baseline generation quality across the evaluated models, offering a favorable trade-off between generation quality and hardware performance.

In summary, this paper makes the following contributions:
\begin{itemize}

\item We identify \textbf{temporal differential sparsity} as a low-cost sparsity signal for diffusion hardware acceleration, and show that only a small fraction of activations need to be recomputed across differential timesteps.

\item {We propose \textbf{\T}, an algorithm--architecture co-designed accelerator, with a light \T~engine for streaming sparsity detection and cache management.}

\item {We develop a \textbf{Condensed Pipeline} that turns these sparse updates into systolic computation across FC, convolution, and attention, without modifying the PE-array dataflow}.

\item We implement \T~and conduct comprehensive evaluations on modern diffusion models, demonstrating significant improvements in performance and memory efficiency compared with prior diffusion accelerators.

\end{itemize}

\section{Background and Motivation}

This section provides the background of diffusion inference and reviews existing diffusion hardware-acceleration techniques. We then present our key observation on temporal differential sparsity, which motivates the \T~architecture.

\begin{figure}[tb]
    \centering
    \includegraphics[width=0.95\linewidth]{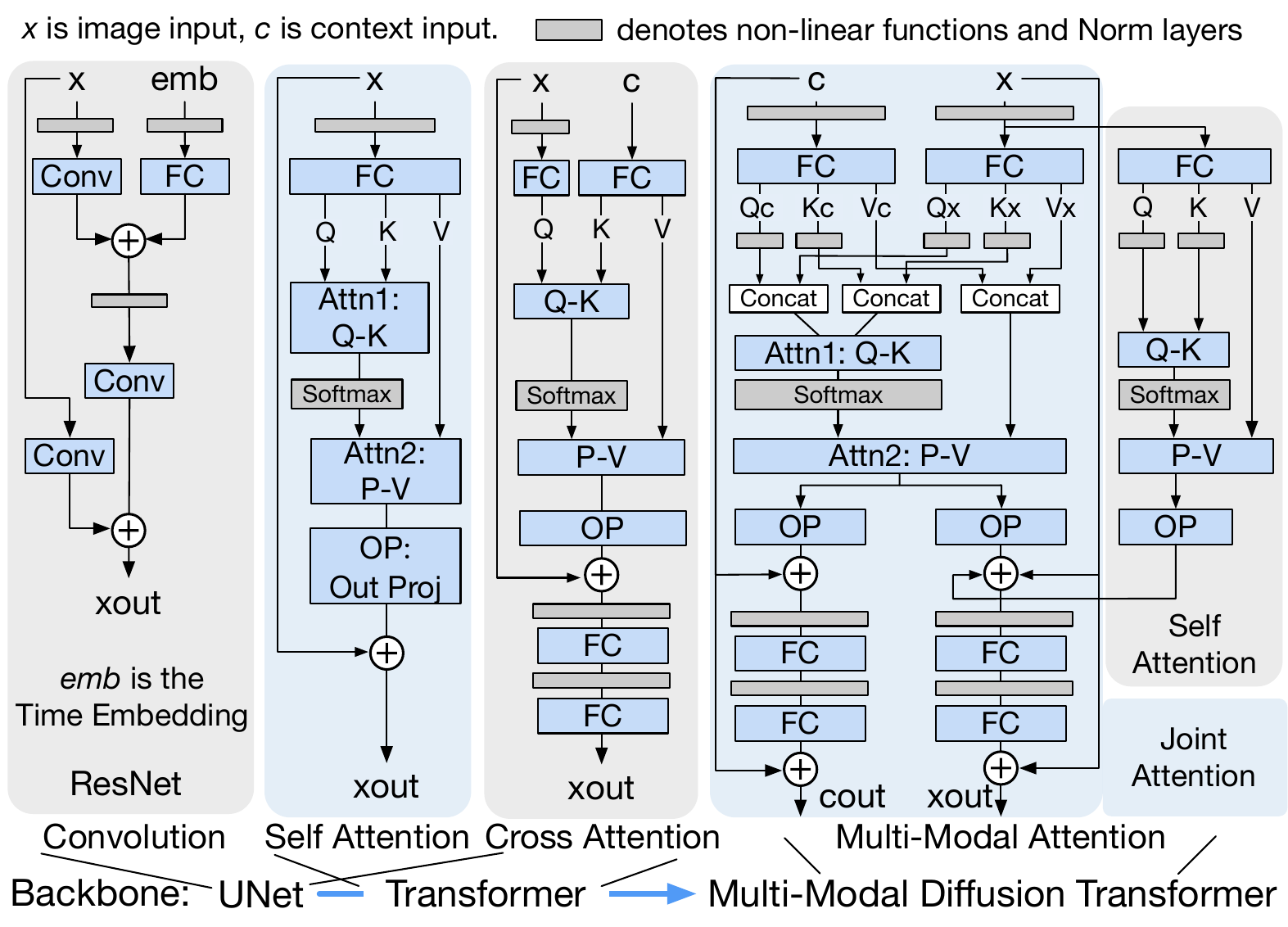}
    \caption{Evolution of diffusion backbones from UNet to Transformer and Multi-Modal DiT.
    }
    \label{fig:bg}
    \vspace{-3mm}
\end{figure}

\subsection{Diffusion Models}
Diffusion models generate images through iterative denoising, where a latent representation is refined across multiple timesteps.
As generation quality improves, diffusion models have evolved toward increasingly complex backbones. As shown in Fig.~\ref{fig:bg}, the backbone has shifted from convolution-heavy UNet structures~\cite{ho2020denoising, song2020denoising, rombach2022high} to attention-based Diffusion Transformers (DiT)~\cite{peebles2023scalable, chen2023pixart}. Recent state-of-the-art models further adopt a Multi-Modal DiT backbone (MMDiT)~\cite{esser2024scaling, labs2025flux1kontextflowmatching, li2026unraveling}. 
While these designs improve generation quality, they significantly increase computational complexity.
During inference, the diffusion network is executed sequentially across multiple timesteps (typically 50--100), requiring full model recomputation at each step and preventing parallelism across timesteps. Meanwhile, the shift toward attention-dominated backbones with larger workloads further amplifies computation and memory demands, leading to high computational cost and long execution latency.

\begin{figure}[t]
    \centering
    \includegraphics[width=\linewidth]{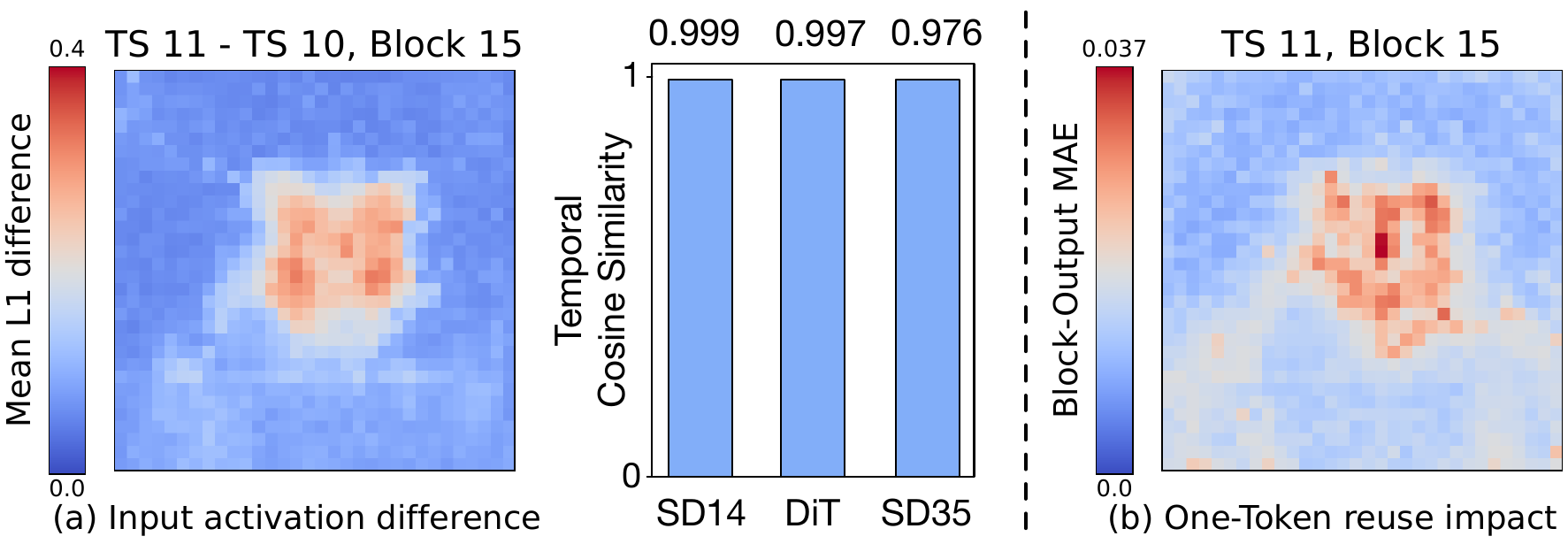}
    \caption{(a) Mean L1 input differences in SD35 (Block 15, steps 10--11; left) and network-wide mean adjacent-timestep activation cosine similarity across models (right). (b) Block-output MAE induced by reusing each token individually.}
    \label{fig:HeapMapAndCosine}
\end{figure}

\subsection{Temporal Similarity across Timesteps}

Although diffusion inference involves many iterations, the activation differences between consecutive timesteps are typically small, which leads to highly similar activations across adjacent timesteps. 
Fig.~\ref{fig:HeapMapAndCosine} (left) shows the mean L1-norm of activation differences, where most differences within the same layer across adjacent timesteps are small, with only a small fraction of outliers changing significantly.  
Moreover, Fig.~\ref{fig:HeapMapAndCosine} (right) presents the cosine similarity between activations of adjacent timesteps. The similarity remains consistently high (e.g., $>0.96$) throughout the diffusion process and across different diffusion models, indicating a strong temporal similarity in diffusion. This strong temporal similarity suggests that consecutive timesteps share highly redundant information, motivating the development of state-of-the-art differential computation techniques.

\subsection{Differential Quantization}

Most prior diffusion accelerators exploit temporal similarity using {differential quantization} (DQ)~\cite{kong2024cambricon,guo202420,tang2025diff,kim2025ditto}, as shown in Fig.~\ref{fig:intro}(b). Instead of full-precision computing, these approaches compute the activation differences between consecutive activations: $\Delta_t = x_t - x_{t+1}$.
Because most differences are small, they can be quantized to low precision, while large outliers are processed in full precision.
Diffy~\cite{mahmoud2018diffy} first introduces DQ for DNN workloads, and CambD~\cite{kong2024cambricon} extends it to UNet-based diffusion models to achieve lower bitwidth and mixed-precision computation (e.g., INT4/FP16). DQ is inherently suited for linear operations (e.g., convolution and FC), but it does not directly apply to attention, where both input matrices in matrix multiplication change, violating the linear assumption of DQ. 
To address this, Ditto~\cite{kim2025ditto} proposes a differential attention mechanism, where a single matrix multiplication is decomposed into two sparse matrix multiplications to account for changes in both inputs. This enables the low-bit optimization for attention under DQ. Based on this, Ditto applies a two-stage quantization scheme (W8A8 + DQ) to represent activations in INT4/8 that supports lower bitwidth than CambD.
While DQ reduces computation cost, it introduces several architectural limitations:

\textbf{Computation overhead.} DQ reduces arithmetic precision but does not inherently reduce the overall computation volume. Moreover, it provides limited optimization for attention, as each matrix multiplication must be decomposed into two matrix multiplications to handle input changes separately~\cite{kim2025ditto}, increasing overall computation.

\textbf{Memory overhead.} While DQ benefits linear layers, it introduces significant data dependencies in non-linear layers requiring full-precision activations (e.g., gray layers in Fig.~\ref{fig:bg}). These dependencies require storing both input and output activations for data conversion between differential and full precision, leading to increased memory footprints.

\textbf{Architectural constraints.}
Quantization-based accelerators aim to improve efficiency by leveraging the smaller area and higher density of low-precision arithmetic units.
However, this benefit is fundamentally limited by the need to co-design heterogeneous compute units for different precisions.
In CambD, although low-bit integer operations (e.g., INT4) could theoretically reuse portions of FP16 units, such reuse cannot fully exploit the area efficiency of dedicated low-precision hardware.
As a result, CambD adopts a mixed-precision design with separate integer and floating-point units to support differential quantization.
This leads to a fixed INT4/FP16 ratio (e.g., 60:4) hard-wired into every PE, breaking the internal homogeneity of the PEs. Furthermore, such a rigid, software-driven mapping reduces architectural flexibility and limits scalability across evolving model requirements.

Ditto partially alleviates the fixed precision constraint through adaptive bitwidth, and its INT4/INT8 mixed design improves hardware utilization by enabling finer-grained resource sharing.
However, it entirely removes FP16 support, restricting the design to integer-only computation.
While this design is efficient for quantization friendly workloads, the lack of floating-point support limits its generality and makes it difficult to extend to more complex or precision-sensitive models~\cite{esser2024scaling, resterched2026flux3}.

\subsection{Sparsity-Based Acceleration}
{
For a sparsity-based accelerator, the benefit of skipping computation depends on both the sparsity available under generation-quality constraints and the cost of identifying and executing irregular work. Ditto~\cite{kim2025ditto} exposes zero-valued element-wise differences by combining W8A8 quantization with differential quantization. Its Encoding Unit classifies and reorders them for zero skipping. However, such sparsity remains tied to quantized differential execution, which incurs substantial quality degradation on SD35 in our evaluation, as shown in Table.~\ref{tab:network_performance}.}
EXION~\cite{heo2025exion} combines two forms of sparsity: it reuses selected FFN outputs across timesteps, whereas attention uses approximate scores to skip low-contribution computations within a timestep. Although EXION is tunable, transformer-based diffusion models are sensitive to such sparsity~\cite{heo2025exion}, requiring conservative settings to limit quality degradation. 
Furthermore, supporting such attention prediction and sparse-output compaction also requires an additional prediction array and triple-buffered PEs in EXION, adding 22\% area overhead. These designs therefore leave unresolved how to exploit temporal redundancy across operators at modest hardware cost while preserving generation quality on modern diffusion models.

\subsection{Differential Sparsity}
Instead of interpreting temporal similarity as quantization and applying recomputation, we observe that it naturally induces \emph{sparsity}. 
Specifically, only a small fraction of activations change significantly between consecutive timesteps, while the majority remain nearly unchanged, as shown in Fig.~\ref{fig:HeapMapAndCosine}. 
We refer to this property as \textbf{differential sparsity}.

Formally, let $\mathbf{x}_t \in \mathbb{R}^{N \times D}$ denote the token-wise activations at timestep $t$, where $N$ is the number of tokens and $D$ is the feature dimension. 
We define the token-wise difference magnitude as:
\begin{equation}
\mathbf{d}_t(i) = \left\lVert \mathbf{x}_t(i) - \mathbf{x}_{t+1}(i) \right\rVert_1, \quad i \in \{1, \dots, N\}
\label{euq1}
\end{equation}
where $\mathbf{x}_t(i)$ denotes the $i$-th token.
We then define the differential mask as the set of token indices corresponding to the top-$k\%$ largest differences:
\begin{equation}
M_t = \text{TopK}_{k\%} \left( \{ \mathbf{d}_t(i) \}_{i=1}^{N} \right)
\end{equation}
where $M_t \subseteq \{1, \dots, N\}$ denotes the indices of outlier tokens.
{As shown in Fig.~\ref{fig:HeapMapAndCosine}(b), tokens with larger input L1 differences tend to induce higher block-output MAE when their previous-timestep activations are reused.}
Under this formulation, only the activations corresponding to tokens in $M_t$ need to be recomputed, while the remaining tokens can be directly reused from the previous timestep via a temporal cache, forming the basis of \emph{differential reuse}. 

This observation motivates a new accelerator architecture for diffusion inference: instead of quantizing temporal differences, we treat them as sparse updates and unify such optimization in all main operations by reusing previous activations through temporal caching . This idea forms the basis of \T's streaming architecture for dynamic cache management and sparse execution, described in the next section.

\section{{\T~Architecture Overview}}
To exploit this observation, we propose \T, an algorithm and architecture co-design framework that turns differential sparsity into hardware speedup, as shown in Fig.~\ref{fig:HW_architecture}.

\subsection{Algorithm-Level Optimization.}
\begin{figure}[t]
    \centering
    \includegraphics[width=0.95\linewidth]{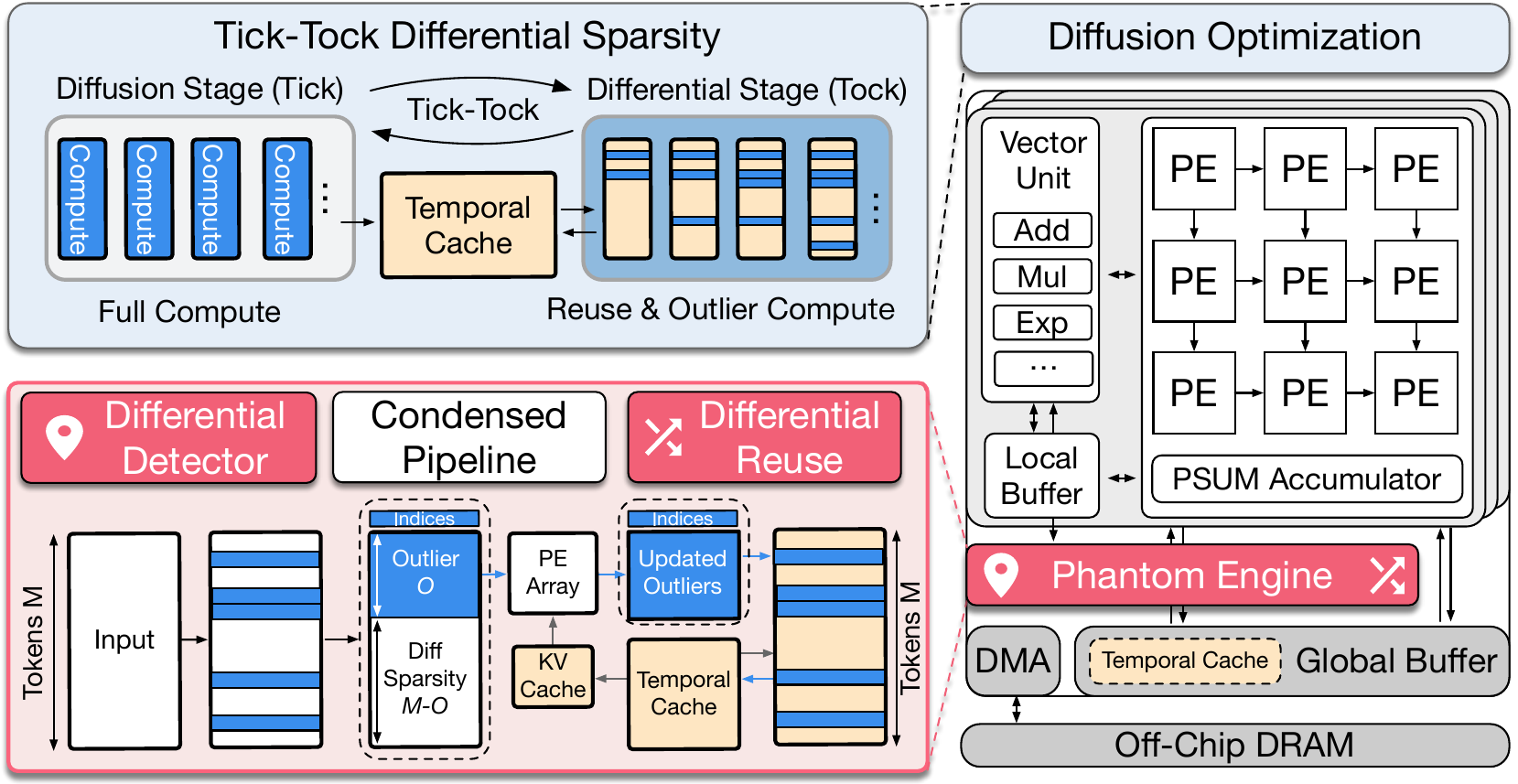}
\caption{{Overview of \T~architecture with software-hardware co-design.
The \T~engine performs dynamic sparsity detection and differential reuse, while the condensed mechanism maps sparse computation to dense execution.}}
    \vspace{-3mm}
    \label{fig:HW_architecture}

\end{figure}

At the algorithm level, \T~pairs differential sparsity with a tick--tock schedule, inspired by the periodic full computation used in caching methods~\cite{zou2024accelerating,liu2025taylorseers} to bound approximation error. 
Timesteps alternate between a \emph{diffusion stage} (tick), which performs standard inference to construct the \textbf{temporal cache}, and a \emph{differential stage} (tock), which detects \textbf{sparsity} and \textbf{reuses} cached activations to skip redundant computation, as shown in Fig.~\ref{fig:HW_architecture} (top left).

Realizing this execution model on hardware is non-trivial. Sparsity must be detected dynamically before each block, and cached activations must be updated at irregular spatial locations. Such data-dependent patterns and irregular accesses are difficult to efficiently support on conventional systolic-array-based accelerators.

\begin{figure*}[t]
    \centering
    \includegraphics[width=0.95\linewidth]{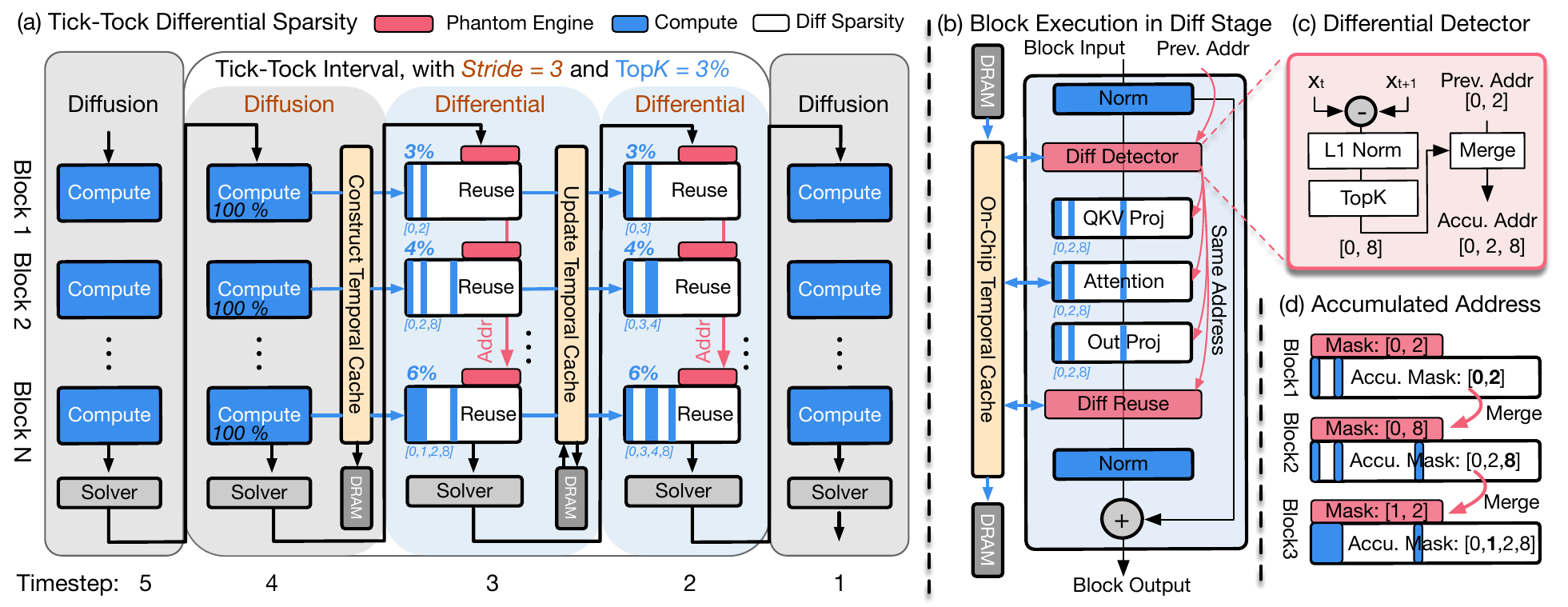}
    \vspace{-3mm}
    \caption{{Overview of the proposed differential sparsity.
(a) Tick–tock strategy with two hyperparameters: stride and top-$K$.
(b) Block execution during the differential stage.
(c) Differential detector design.
(d) Accumulated mask mechanism. Diff denotes differential and Accu. denotes accumulated.}}
    \label{fig:adaptive_timestep}
    \vspace{-3mm}
\end{figure*}

\subsection{Architecture-Level Optimization.}
\label{sec:overview}
\T~builds on a conventional hierarchical accelerator, as shown in Fig.~\ref{fig:HW_architecture} (right): multiple systolic arrays, each with local buffers dedicated to GEMM tiling, share a global buffer that stages activations and weights from off-chip memory. This architecture provides high parallelism and fine-grained tiling flexibility across the diverse layer shapes in diffusion workloads. To accelerate the diffusion models above without sacrificing these flexibilities,
\T~introduces a modular \T~Engine for dynamic sparsity handling, together with a Condensed Pipeline for efficient sparse execution.

\textbf{\T~Engine} comprises a differential detector for streaming sparsity detection and a differential reuse unit for efficient activation reuse. It is attached to the global buffer rather than to any individual array: detection and reuse operate on tiles passing through the global buffer, where the temporal cache also resides, and condensed tiles are then delivered to the arrays along the standard tiling path. Consequently, the computing dataflow inside each array remains unmodified, and a single Engine serves all arrays, amortizing rather than replicating its cost.

\textbf{Condensed Pipeline.}
Differential sparsity induces row-sparse activation patterns, which prior work~\cite{kim2025ditto} does not fully exploit in attention. The condensed pipeline introduces a lightweight KV cache to reconstruct full-size $K$ and $V$ tensors (Fig.~\ref{fig:HW_architecture} (bottom left)), so that row-sparse computation executes as dense attention with high utilization.

We detail the tick-tock differential sparsity in Section~\ref{sec:HTS}, the \T~Engine in Section~\ref{sec:Engine}, and the condensed pipeline in Section~\ref{sec:Condense}.

\section{Tick-Tock Differential Sparsity}
\label{sec:HTS}
To ensure stable generation with the differential sparsity, we use a Tick-Tock strategy (Fig.~\ref{fig:adaptive_timestep}) for differential sparsity. It contains two key components: (1) temporal cache construction from a diffusion stage, and (2) sparsity generation via a differential stage.

\subsection{Overview}
The diffusion stage serves as a pre-processing step, which performs standard diffusion inference to construct and fully refresh the temporal cache. The differential stage (tock) compares current block inputs with their cached values to detect sparsity, recomputes selected tokens, and updates the corresponding cache entries while reusing the remaining cached results (Fig.~\ref{fig:adaptive_timestep}(a)). However, if all timesteps adopt the differential stage, the resulting sparsity becomes increasingly localized, leaving some regions persistently inactive and never updated. This issue can be naturally addressed by the diffusion stage, as the full computation restores global information propagation.

Based on this observation, our method alternately executes timesteps between these two stages: a diffusion stage (tick) followed by a differential stage (tock), forming a tick–tock interval.
The number of timesteps per interval is controlled by a hyperparameter, \textit{stride}. When $stride = s$, each interval consists of one standard timestep in the diffusion stage and $s-1$ timesteps in the differential stage (e.g., $stride=3$ in Fig.~\ref{fig:adaptive_timestep}(a)). To preserve generation quality, the first and last timesteps of diffusion inference are always assigned to the diffusion stage, while the remaining timesteps follow the tick–tock strategy.

\subsection{Diffusion Stage}

\textbf{Temporal Cache.}
As a pre-processing step in the tick-tock interval, the diffusion stage performs one standard diffusion timestep to construct a precise temporal cache for subsequent differential reuse, as shown in Fig.~\ref{fig:adaptive_timestep}(a).
To limit the memory overhead, only activations involved in sparsity-related operations are cached: block inputs for differential sparsity detection, outputs of the last FC/MLP/Conv layer in each block for differential reuse, and selected intermediate results (e.g., attention KV tensors and convolution inputs) for the condensed pipeline (Section~\ref{sec:Condense}).

{The temporal cache of all layers is written to DRAM during diffusion steps and pre-fetched back layer-by-layer during differential steps, so only the tensors of the active layer reside on-chip. The on-chip organization of temporal cache is detailed in Section~\ref{sec:Engine}.}
Since the diffusion stage is periodically invoked in the tick–tock strategy, the temporal cache is refreshed via full computation, preventing error accumulation from differential sparsity and restoring global information propagation.

\begin{figure*}[tb]
    \centering
    \includegraphics[width=0.95\linewidth]{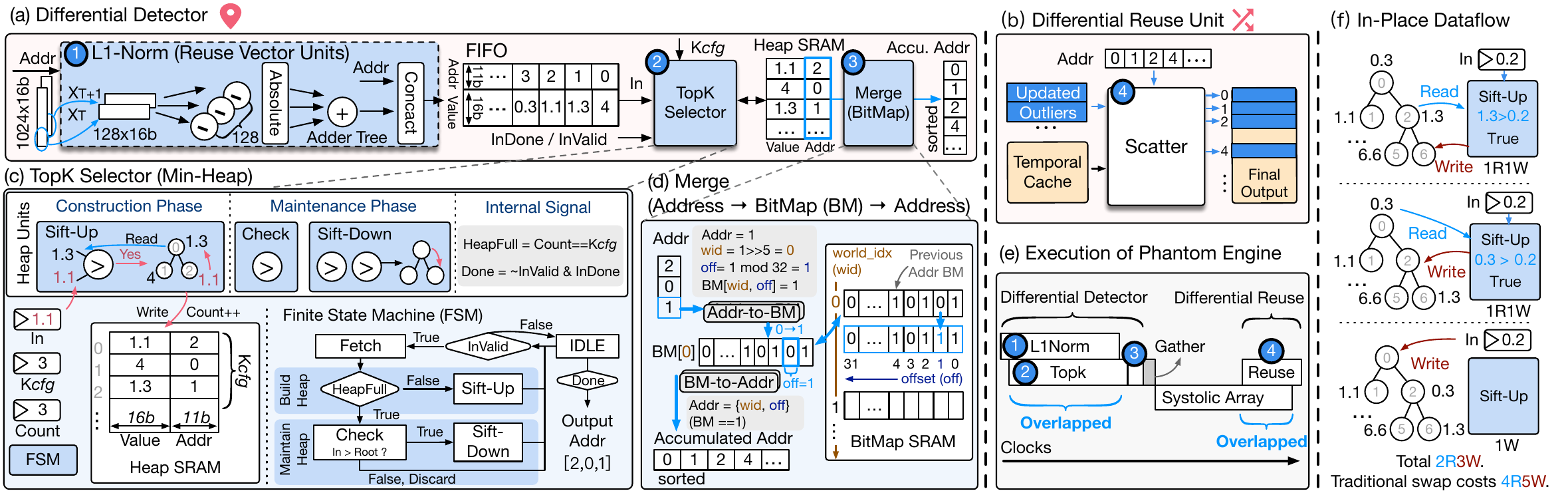}
    \caption{Hardware design of the \T~Engine. It comprises (a) a differential detector and (b) a differential reuse unit. The detector integrates (c) a min-heap-based TopK selector and (d) a bitmap-based merge unit to enable efficient outlier selection and address merging. (e) presents the pipelined execution flow. (f) Sift-up heap construction example with in-place dataflow to reduce memory accesses. \textit{cfg} denotes configuration signals. \textit{Addr} denotes Address.}
    \label{fig:PhantonEngine}
    \vspace{-3mm}
\end{figure*}

\subsection{Differential Stage}

The differential stage accelerates diffusion inference by exploiting differential sparsity. It contains multiple timesteps in the tick-tock interval, where sparsity is detected at each block's input for each timestep. During the differential stage, only highly-changed tokens are computed, while the remaining tokens (differential sparsity) reuse cached results to avoid redundant computation. Fig.~\ref{fig:adaptive_timestep}(b) illustrates the block-level execution flow of the differential stage. 

\textbf{Differential sparsity} is first identified by a \textit{differential detector}. When block execution reaches timestep $t$, its incoming input is $\mathbf{x}_t$, while the temporal cache still holds the same block's input from the previous timestep, i.e., $\mathbf{x}_{t+1}$ in Equation~(\ref{euq1}). The detector therefore evaluates Equation~(\ref{euq1}) directly between the incoming block input and its cached copy, as shown in Fig.~\ref{fig:adaptive_timestep}(c).
Sparsity is controlled by a hyperparameter $K$, where the Top-$K$ tokens with the largest changes are selected as \textit{outliers}, and the rest are treated as sparse. Since $K$ is typically small (5\%–15\%), \T~records the outlier addresses for subsequent computation. 
In practice, different blocks may exhibit dynamic computation focus, causing a potential shift of outliers across blocks. Independent detection at each block can therefore discard outliers identified earlier, leading to lost updates. To address this, the current outlier addresses are incrementally merged with those from previous blocks, forming an accumulated address set that ensures consistent updates, as shown in Fig.~\ref{fig:adaptive_timestep}(d). Consequently, subsequent computations are performed only on these outliers. Accumulated masks reset at each timestep boundary, while temporal caches persist across timesteps.

\textbf{Differential Reuse.} After all computation layers (Attention / FC / Conv) in the block are completed, the differential reuse fills in the sparse entries using the temporal cache to reconstruct the full tensor. The final normalization and residual operations are then applied. The updated outlier activations are written back to the cache to keep it up to date for subsequent timesteps.
Note that differential sparsity and reuse are applied only to image tokens, while text tokens are always fully computed, as they contribute relatively little to computation but play a critical role in guiding generation quality.
For UNet-based diffusion models, we do not apply address accumulation across blocks, as outliers already exhibit high spatial overlap (95\%), and token addresses are not directly reusable across blocks due to inconsistent spatial correspondence.

\textbf{Summary.}
The tick-tock differential sparsity provides token-level dynamic sparsity control within the differential stage, tunable through the stride and Top-\textit{K} hyperparameters to match different diffusion models and quality requirements.
Compared to differential quantization, which relies on heavily customized designs, the flexibility from differential sparsity ensures both generality and effectiveness in our hardware architecture. The hardware challenges of dynamic sparsity detection and sparse skipping are addressed in the following sections.

\section{\T~Engine}
\label{sec:Engine}
{As introduced in Section~\ref{sec:overview}, the \T~Engine operates at the global buffer and consists of two units: a \textit{differential detector} that identifies outlier tokens before each block, and a \textit{differential reuse} unit that fills in the remaining tokens from the temporal cache after computation. 
We propose a streaming design for them to minimize the impact of sparsity handling on the critical path, with reuse fully overlapped by systolic-array execution. Since both operate on the temporal cache that resides in the global buffer, we first describe how this cache is organized on-chip, then detail the two units.}

\begin{figure*}[tb]
    \centering
    \includegraphics[width=0.95\linewidth]{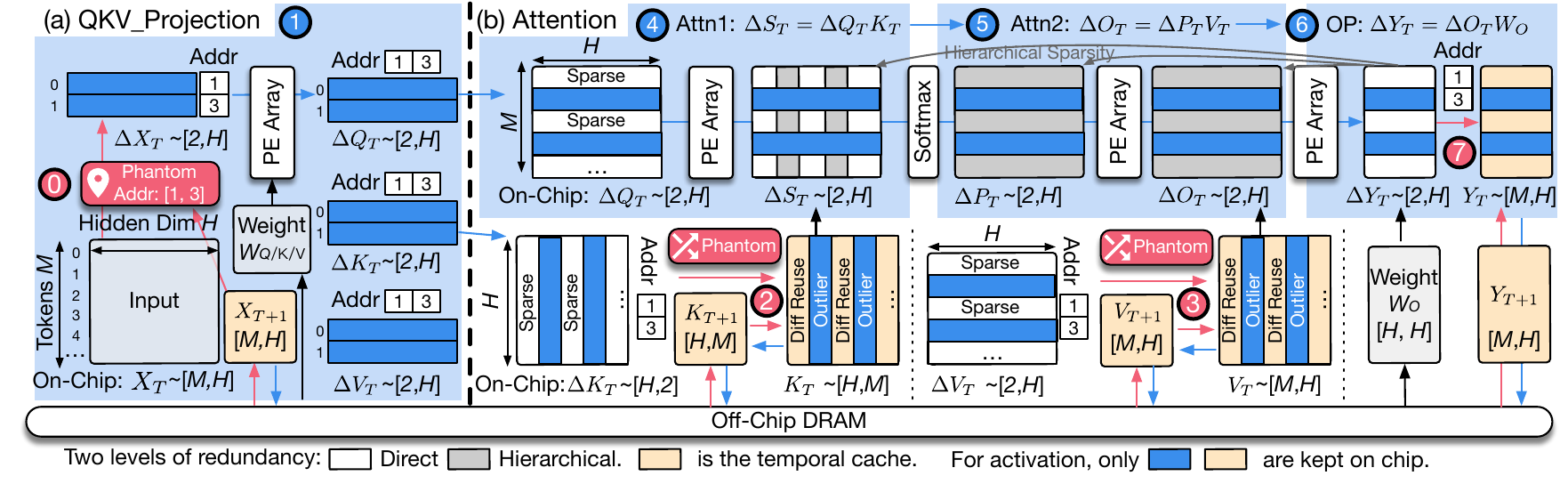}
    \vspace{-2mm}
    \caption{Condensed pipeline for (a) QKV projection (FC layers) and (b) attention layers. 
OP denotes the output projection. 
$\Delta X$ represents the outlier set of matrix $X$, $[M,N]$ is the matrix dimensions. $T$ and $T\!+\!1$ denote the current and previous timesteps.}
    \label{fig:condensed}
\end{figure*}

\textbf{Temporal Cache.} {The temporal cache is not a dedicated structure but a \textit{stage-dependent partition} of the global buffer. During the differential stage, a partition of the global SRAM 
is allocated as the temporal cache and organized as a double buffer over detection and reuse tiles, pipelining cache reads and write-backs with PE-array computation. During the diffusion stage the partition disappears, as full computation writes cache updates through the ordinary output SRAM path. The temporal cache therefore consumes no dedicated area and materializes only when the differential stage requires it.}

\subsection{Differential Detector}
The differential detector performs three steps: token-wise L1-norm computation, online TopK selection, and bitmap-based address accumulation and sorting, as shown in Fig.~\ref{fig:PhantonEngine}(a). The first step reuses existing hardware, the L1-norm between the current activation $X_T$ and its cached counterpart $X_{T+1}$ is a per-token reduction mapped onto the on-chip vector unit, and its \{value, address\} outputs are streamed into a FIFO inside the Engine. The remaining two steps use dedicated units: a configurable TopK selector consumes these pairs online to identify outlier tokens, and a merge unit accumulates their addresses with previously selected ones into a sorted list, enabling regular memory access for subsequent operations.

\textbf{TopK Selector} is optimized with an SRAM-based min-heap, as shown in Fig.~\ref{fig:PhantonEngine}(c). Heap storage is allocated from the shared global SRAM pool. The heap is dynamically configured by $K_{cfg}$ and operates in two phases under a fine-grained FSM. In the construction phase, incoming entries are inserted at the last node and the heap is built by the \textit{Sift-Up} unit. Once the heap reaches $K_{cfg}$ entries, it enters the maintenance phase, where each new input is first compared with the root (i.e., the smallest of the maintained entries) by the \textit{Check} unit: smaller inputs are discarded immediately, while larger ones replace the root and trigger the \textit{Sift-Down} unit. This root comparison discards a large fraction of incoming entries directly (around 53.0\% on SD35), so that the selector tracks outliers in a streaming manner and its latency is overlapped with L1-norm computation, as shown in Fig.~\ref{fig:PhantonEngine}(e). An in-place update dataflow further shifts elements along the update path and writes the input entry only once at its final position, saving around 50\% of heap SRAM accesses compared with conventional swap-based updates (Fig.~\ref{fig:PhantonEngine}(f)).

\textbf{Bitmap-Based Address Merge.} The merge unit is implemented on a bitmap representation, which naturally supports both merging and sorting. As shown in Fig.~\ref{fig:PhantonEngine}(d), each incoming address is translated into a bitmap word index (wid) and a bit offset (off) through \textit{Addr-to-BM} units. A 32-bit bitmap word encodes the outlier status of the corresponding tokens, and newly detected addresses are merged via a read-modify-write update using a bitwise OR. Accumulation across blocks is supported by simply maintaining the bitmap. After merging, \textit{BM-to-Addr} units scan the bitmap SRAM and reconstruct the address list, inherently producing sorted addresses that benefit both the gather stage of condensed computation and the differential reuse by avoiding irregular memory accesses.

\subsection{Differential Reuse Unit}
The differential reuse unit is lightweight and performs scatter-style write-back based on the sorted outlier addresses. As shown in Fig.~\ref{fig:PhantonEngine}(b), updated outlier activations are written back to the temporal cache to reconstruct the full output tensor and refresh cached activations for subsequent timesteps. Since the addresses are already sorted, the reuse process is streamed with regular memory access and overlapped with PE-array execution, allowing outliers to be written back immediately after their partial computations complete.

\textbf{Summary.} 
The \T~Engine is naturally tile-decomposable: detection performs a per-token reduction and reuse a per-token scatter, so tiles are processed as they stream through the global buffer. Differential reuse is designed to be fully overlapped with systolic-array execution, while detection cannot, as it depends on the entire block input. After its internal pipeline optimization, detection only costs a 0.7\% of block-level critical-path latency. With the L1-norm mapped onto the existing vector unit, the Engine adds only lightweight logic and marginal area.

\begin{figure}[tb]
    \centering
    \includegraphics[width=0.9\linewidth]{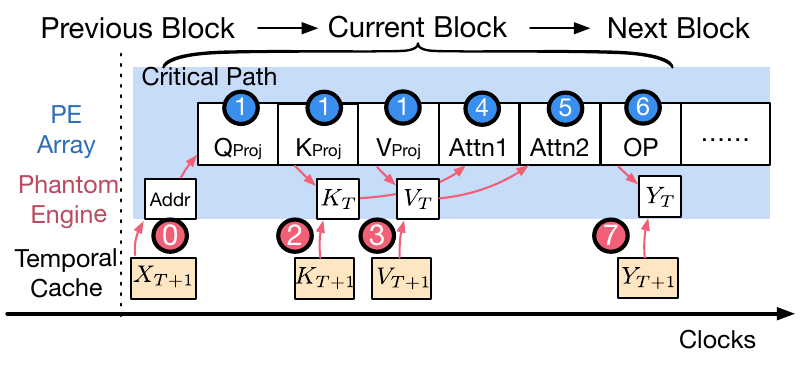}
    \vspace{-3mm}
    \caption{Detailed dataflow of the condensed attention with overlapped KV reconstruction in the differential stage.}
    \label{fig:schedule}
\end{figure}

\section{Condensed Pipeline}
\label{sec:Condense}
While the \T~Engine detects and reuses differential sparsity, the resulting row-sparse execution remains underexplored. We therefore propose a \textit{Condensed Pipeline}, realized within the \T~Engine, which compacts sparse activations into dense inputs, enabling dense execution of fully connected (FC), convolution (Conv), and attention layers without modifying the core systolic-array dataflow.

\subsection{Condensed FC/Convolution}
The row-sparse pattern from differential sparsity can be easily exploited in linear operators such as FC, Conv, and token-wise normalization layers.
Fig.~\ref{fig:condensed}(a) illustrates the QKV projection prior to attention. After the \T~Engine identifies outlier addresses \circlednumberred{0}, the memory control unit gathers only the corresponding tokens ($\Delta X_T$, where $T$ denotes the current timestep) while preserving their original addresses, which naturally form dense inputs for \circlednumberblue{1} FC computation. 
Conv requires a full input receptive field for each selected output position. {\T~stores the current convolution inputs after preprocessing and gathers the required patches for Conv-to-FC execution, after which the condensed tokens follow the same dense flow as FC.
In UNet, pre-convolution normalization and timestep conditioning are fully recomputed, while convolution and post-convolution token-wise normalization run only at selected output positions.}

\subsection{Condensed Attention}
When outlier tokens are updated, both operands ($\Delta Q_T$ and $\Delta K_T/\Delta V_T$) of the attention matrix multiplications change, hindering direct exploitation of sparsity, as in prior differential approaches~\cite{kong2024cambricon,kim2025ditto,tang2025diff}. Differential attention~\cite{kim2025ditto} decomposes it into two sparse matrix multiplications that separately update the affected regions (the blue and gray regions of $\Delta S_T$ in Fig.~\ref{fig:condensed}(b)), but this introduces dependence on attention scores and may exceed standard attention latency when sparsity is low. \textit{Condensed Attention} instead condenses outlier tokens into dense operands to skip computation directly, and proactively caches $K$ and $V$ to reconstruct full-size operands without depending on previous attention scores.

\textbf{Two levels of redundancy} are exploited in the condensed attention. The first is the direct redundancy from the row-sparse outputs of the QKV projection layer, where computations for non-outlier tokens are skipped, corresponding to the white regions in Fig.~\ref{fig:condensed}(b)\circlednumberblue{4}. The second is the hierarchical redundancy from the output projection (OP) layer that follows attention: as an FC layer, \circlednumberblue{6} OP also operates only on outlier rows, which propagates backward to compute selected-query attention over reconstructed K/V, allowing the gray regions of \circlednumberblue{4} Attn1 and \circlednumberblue{5} Attn2 to be skipped (Fig.~\ref{fig:condensed}(b)). Together, the two levels reduce attention to a dense computation over the outlier set.

\textbf{KV cache} is proactively introduced to realize this execution. During the previous diffusion stage, the key and value matrices are cached as $K_{T+1}$ and $V_{T+1}$. In the differential stage, reconstruction is fused into the KV projection rather than forming a separate step: as the PE array produces each tile of $\Delta K_T$, the differential reuse unit merges it into the cached $K_{T+1}$ at the corresponding outlier addresses, so that the full-size $K_T$ is ready once the projection completes (Fig.~\ref{fig:condensed}(b)\circlednumberred{2}\circlednumberred{3}). $V_T$ is updated in the same way. When capacity permits, the reconstructed tensors remain in the temporal cache for direct reuse; otherwise they are streamed through DRAM at tile granularity. 
Since the reuse unit writes each projection tile while the PE array computes the next, the merge is fully overlapped with the projection and adds no critical-path latency (Fig.~\ref{fig:schedule}).

The outlier query $\Delta Q_T$ is then multiplied with $K_T$ tile by tile to update the outlier rows of the attention scores $\Delta S_T$, which proceed through softmax. Subsequent row-sparse computations ($\Delta P_T V_T$ and $\Delta O_T W_O$) follow the same flow, after which differential reuse reconstructs the output $Y_T$. 
Moreover, unlike the LLM KV cache that grows with sequence length, the KV cache in diffusion is bounded by the fixed token count of the generated image, so its size stays constant across timesteps and its memory overhead remains lightweight.

\textbf{Summary.} With the condensed pipeline, the row-sparse execution induced by differential sparsity is converted into proportional reductions in computation latency and memory traffic on an unmodified systolic array, closing the gap between exposed sparsity and realized speedup.

\begin{table}[tb]
\centering
\small
\caption{Evaluated Networks and Datasets.}
\scalebox{0.8}{
\begin{tabular}{lcccc}
\toprule[1.5pt]

\textbf{Notation}
& \textbf{Model}
& \textbf{Backbone}
& \makecell{\textbf{Dataset} /\\ \textbf{Resolution}}
& \makecell{\textbf{Sampler} /\\ \textbf{Steps}} \\

\midrule[1.5pt]

SD14
& {\makecell{Stable Diffusion\\1.4~\cite{rombach2022high}}}
& UNet
& \makecell{MSCOCO~\cite{lin2014microsoft} \\ 256, 512}
& \makecell{PLMS~\cite{liu2022pseudo} \\ 50 steps} \\

\midrule[0.5pt]

DiT
& {\makecell{DiT-XL/2~\cite{peebles2023scalable}}}
& Transformer
& \makecell{ImageNet~\cite{deng2009imagenet} \\ 256, 512}
& \makecell{DDIM~\cite{song2020denoising} \\ 100 steps} \\

\midrule[0.5pt]

SD35
& {\makecell{Stable Diffusion\\3.5~\cite{esser2024scaling}}}
& \makecell{MMDiT\\Transformer}
& \makecell{MSCOCO~\cite{lin2014microsoft} \\ 512, 1024, 1536}
& \makecell{ODE~\cite{lipman2022flow} \\ 50 steps} \\

\bottomrule[1.5pt]
\end{tabular}
}
\label{tab:evaluatedModels}
\end{table}

\section{Evaluation}

\subsection{Methodology}

\textbf{Models and Workloads.}
We evaluate \T~on three representative diffusion models (Table~\ref{tab:evaluatedModels}): Stable Diffusion 1.4 (SD14)~\cite{rombach2022high}, Diffusion Transformer (DiT)~\cite{peebles2023scalable}, and Stable Diffusion 3.5 (SD35)~\cite{esser2024scaling}, spanning the major state-of-the-art (SOTA) diffusion backbones and covering both text-to-image and class-conditional generation.
Following standard protocols, we generate 5K images at 512$\times$512 on MSCOCO~\cite{lin2014microsoft} for SD14 and SD35 (50 timesteps), and 50K images at 256$\times$256 on ImageNet~\cite{deng2009imagenet} for DiT (100 timesteps).
Generation quality is measured by FID~\cite{heusel2017gans}, Inception Score (IS)~\cite{salimans2016improved}, and CLIP Score (CS)~\cite{hessel2021clipscore}.
All models use officially released checkpoints from HuggingFace~\cite{huggingface}, implemented with the Diffusers framework~\cite{von-platen-etal-2022-diffusers} on PyTorch~\cite{paszke2019pytorch} with their default model configurations for faithful reproduction.
To further evaluate hardware scalability, we benchmark all designs across generation resolutions from 256$\times$256 to 1536$\times$1536 for each model, spanning workloads with substantially different compute and memory demands.

\begin{table}[tb]
\centering
\small
\caption{Evaluated Hardware Accelerators.}
\setlength{\tabcolsep}{3pt}
\renewcommand{\arraystretch}{1.15}
\resizebox{\columnwidth}{!}{
\begin{tabular}{lccccc}
\toprule[1.5pt]
 & \textbf{TPU} & \textbf{CambD}~\cite{kong2024cambricon} & \textbf{Ditto}~\cite{kim2025ditto} & \textbf{EXION}~\cite{heo2025exion} & \textbf{\textit{\T}} \\
\toprule[1.5pt]
PE Array & $16{\times}16{\times}16$ & $16{\times}16{\times}64$ & $16{\times}16{\times}16$ & $16{\times}16{\times}16$ & $16{\times}16{\times}16$ \\
Array Number & 16 & 4 & 16 & 16 & 16 \\
Act. Bit & FP16 & INT4/FP16 & INT4/8 & FP16 & FP16 \\
Wgt. Bit & FP16 & FP16 & INT8 & FP16 & FP16 \\
SRAM (MB) & 20 & 20 & 20 & 20 & 20 \\
\midrule
Shared Config. & \multicolumn{5}{c}{1\,GHz, GDDR6 @ 819\,GB/s, 28\,nm} \\
\midrule
Logic Area (mm$^2$) & 56.47 & 45.24 & 45.8 & 77.27 & 56.59 \\
Total Area (mm$^2$) & 88.57 & 78.95 & 77.91 & 107.8 & 88.69 \\
\toprule[1.5pt]
\end{tabular}}
\label{tab:hardwareConfigs}
\vspace{-3mm}
\end{table}

\textbf{Baselines.}
We compare \T~against five baselines spanning two categories: one GPU-based baseline and four hardware accelerators.
The GPU baseline is an NVIDIA A100 with FlashAttention~\cite{dao2022flashattention}.
The hardware accelerators category comprises Cambricon-D (CambD)~\cite{kong2024cambricon}, Ditto~\cite{kim2025ditto}, EXION~\cite{heo2025exion}, and a systolic-array TPU with the same hardware budget as \T.
We follow the original methodologies and available implementations of the competing designs without retraining or fine-tuning.

{\textbf{Baseline configurations and operating points.}}
{For the main evaluation, we prioritize FID, one of the most widely used metrics for generated images, and select the fastest \T~configuration that retains near-baseline FID. The stride and TopK of \T~are selected through the design-space exploration in Fig.~\ref{fig:Stride_and_Topk}. The selected configuration is used consistently for quality and performance evaluation.
CambD adopts an INT4/FP16 configuration to better preserve generation quality. 
Ditto uses its original W8A8 framework, on top of which DQ further reduces activations to mixed INT0/4/8 precision.
Note that DQ in CambD and Ditto reduces bitwidth only during computation, while memory accesses remain at their original precision (FP16 for CambD, INT8 for Ditto).
EXION is evaluated in FP16, using its default configuration for SD14. For DiT and SD35, we use ($N{=}1$, FFN sparsity ${=}0.5$, $k{=}0.8$, $q_{\mathrm{th}}{=}1.0$) and ($N{=}2$, FFN sparsity ${=}0.8$, $k{=}0.5$, $q_{\mathrm{th}}{=}0.6$), respectively, selected from a parameter sweep to retain as much sparsity as possible while matching a similar generation quality of the other designs.}

\textbf{Hardware Implementation.}
{For fair comparison, all designs are unified under the same compute and memory budget: 16 independent 16$\times$16 systolic PE arrays with 16 MAC lanes per PE, similar to SOTA works~\cite{heo2025exion, kong2024cambricon}, along with identical off-chip DRAM bandwidth and 20\,MB of on-chip SRAM, whose internal organization follows each design. In \T, this SRAM is organized as 640\,KB of local buffer per array, 
and a 10\,MB global buffer shared by all arrays.
Beyond the unified PE arrays, design-specific units follow each work's original designs, such as Ditto's reconfigurable PEs and EXION's Eager Prediction and Vector Merging units. CambD's PEs are provisioned with 64 MAC lanes (60 INT4 + 4 FP16) to meet its algorithmic requirements.}
All designs are implemented in SystemVerilog and synthesized with Synopsys Design Compiler in TSMC 28nm at 1GHz. SRAM area and energy are modeled with CACTI~\cite{muralimanohar2009cacti} and DRAM energy with DRAMSim3~\cite{li2020dramsim3}.
To evaluate architectural performance, we develop a cycle-accurate simulator based on the open-source framework~\cite{wei2026focus}, which models cycle counts and memory accesses using real network data and is validated against RTL simulation. For fair comparisons with the A100 in Fig.~\ref{fig:speedup_power_area}(b), power is scaled to 7nm using DeepScaleTool~\cite{sarangi2021deepscaletool}.
GPU latency is measured with synchronized timing after warmup, and GPU power via NVML~\cite{nvml} sampling averaged over each workload. 
Detailed hardware configurations are listed in Table~\ref{tab:hardwareConfigs}.

\subsection{Model Accuracy}

\begin{table}[t]
\centering
\caption{Generation quality comparison.}
\label{tab:network_performance}
\footnotesize
\setlength{\tabcolsep}{3pt}
\renewcommand{\arraystretch}{1.15}
\newcolumntype{C}{>{\centering\arraybackslash}X}
\begin{tabularx}{\columnwidth}{@{\hspace{4pt}}>{\raggedright\arraybackslash}p{45pt}|CC|CC|CC@{\hspace{4pt}}}
\toprule
\multicolumn{1}{c|}{\multirow{2}{*}{\raisebox{-0.5ex}{Method}}}
& \multicolumn{2}{c|}{SD14~\cite{rombach2022high}}
& \multicolumn{2}{c|}{DiT~\cite{peebles2023scalable}}
& \multicolumn{2}{c}{SD35~\cite{esser2024scaling}} \\
& FID↓ & CS↑
& FID↓ & IS↑
& FID↓ & CS↑ \\
\midrule
Baseline
& 24.53 & 0.78
& 6.12 & 275.0
& 30.19 & 0.80 \\
CambD~\cite{kong2024cambricon}
& 25.66 & 0.78
& 6.33 & 259.9
& 32.81 & 0.79 \\
Ditto~\cite{kim2025ditto}
& 26.19 & 0.78
& {6.54} & {217.4}
& 47.56 & 0.76 \\
{EXION}~\cite{heo2025exion}
& 24.72 & 0.78
& 6.98 & \textbf{{267.7}}
& {30.5} & {0.79} \\
\midrule
\T
& \textbf{24.46} & \textbf{0.78}
& \textbf{6.23} & 265.5
& \textbf{29.74} & \textbf{0.79} \\
\emph{Config.}
& \multicolumn{2}{c|}{[9, 10\%]}
& \multicolumn{2}{c|}{[6, 5\%]}
& \multicolumn{2}{c}{[6, 15\%]} \\
\bottomrule
\end{tabularx}
\end{table}

\begin{figure}[tb]
    \centering
    \includegraphics[width=\linewidth]{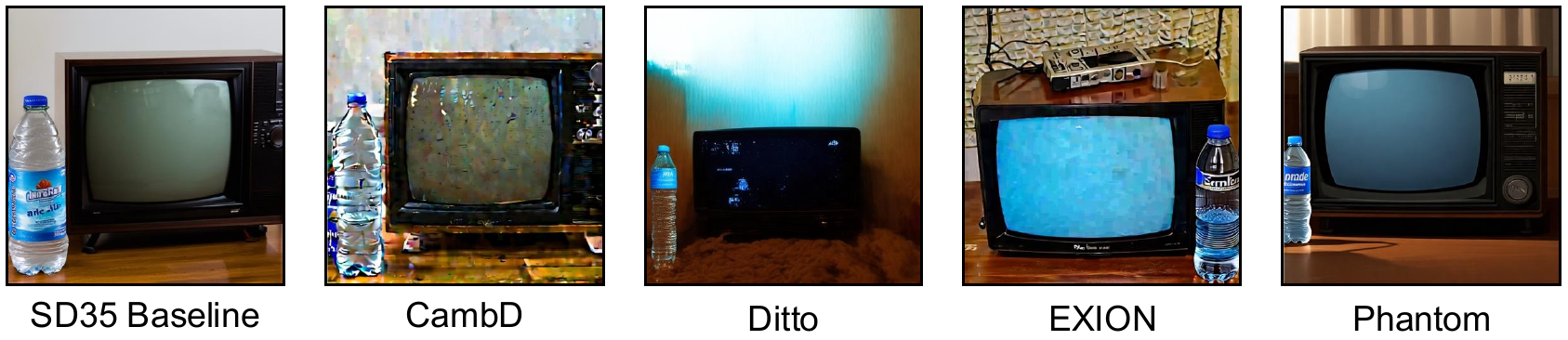}
    \vspace{-5mm}
    \caption{{Image comparison among different works on SD35.}}
    \label{fig:imageCompare}
    \vspace{-3mm}
\end{figure}

\begin{figure*}[tb]
    \centering
    \includegraphics[width=0.95\linewidth]{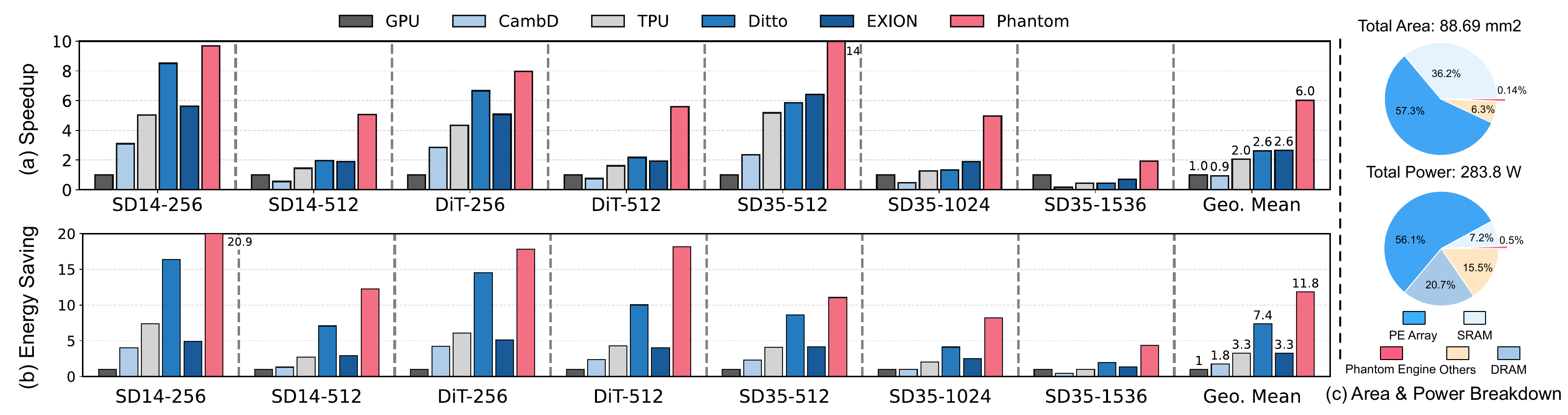}
    \vspace{-3mm}
    \caption{{Comparison results in (a) speedup and (b) energy saving, with GPU as the baseline. Higher speedup and energy saving indicate better performance.} (c) Area and power breakdown.}
    \vspace{-3mm}
    \label{fig:speedup_power_area}
\end{figure*}

\textbf{Model Accuracy.}
Table~\ref{tab:network_performance} reports generation quality at the same operating points used in Fig.~\ref{fig:speedup_power_area}. {Across all three models, \T's FID deviates by no more than 1.8\% from the dense baseline. It is lower than CambD, Ditto, and EXION on each model. Together with Fig.~\ref{fig:speedup_power_area}, these results show that \T~achieves higher average hardware speedup and energy savings while maintaining near-baseline FID.
Fig.~\ref{fig:imageCompare} further shows representative SD35 outputs of different works.}

\subsection{Performance and Energy Evaluation}
We compare \T~with GPU, TPU, CambD, Ditto, and EXION in Fig.~\ref{fig:speedup_power_area} in terms of speedup and energy saving, using GPU as the baseline.

\textbf{Performance.}
\T~achieves the highest speedup as shown in Fig.~\ref{fig:speedup_power_area}(a), averaging {4.0$\times$ over GPUs and SOTA accelerators}. 
Compared with the vanilla TPU processing dense data, \T~delivers a 2.95$\times$ average speedup: differential sparsity exposes an average of 67.8\% sparsity across diffusion workloads, and the condensed pipeline converts it into proportional speedup on the PE array.

Compared with differential quantization designs, \T~achieves 6.38$\times$ speedup over CambD and 2.30$\times$ over Ditto. 
CambD benefits mainly from lower-precision arithmetic, whereas Ditto further improves throughput through zero skipping and reconfigurable PEs. However, they both incur data-conversion overheads around non-linear functions, and Ditto's differential attention can be slower than standard attention at low sparsity.
In contrast, \T~directly skips computation via differential sparsity, consistently accelerating all operations without precision switching.

Compared with EXION, \T~achieves a {2.28$\times$} average speedup under the quality-aware configurations used in our evaluation. On DiT, more aggressive EXION settings substantially degrade generation quality, motivating the less aggressive configuration used for both quality and performance comparisons. Moreover, EXION's sparsity optimizations leave convolutions untouched, whereas \T~applies differential sparsity across attention, FFN, and convolution layers, achieving both broader coverage and higher exploited sparsity.

Compared with GPUs, \T~achieves 6.03$\times$ speedup over the GPU with FlashAttention, owing to dedicated datapaths that sustain high PE utilization on fine-grained sparse computation, which GPUs cannot efficiently exploit.

\textbf{Energy Efficiency.}
Fig.~\ref{fig:speedup_power_area}(b) reports the energy saving of all designs, normalized to the GPU.
On average, \T~consumes 11.8$\times$, 6.74$\times$, 3.61$\times$, 1.61$\times$, and {3.63$\times$} less energy than the GPU, CambD, TPU, Ditto, and EXION, respectively.
These savings come from both computation and memory: by exploiting differential sparsity, \T~reduces the computation volume and the effective activation footprint simultaneously, saving on-chip SRAM accesses, off-chip DRAM traffic, and latency at once. 
\T~also avoids the precision-conversion and additional attention computation incurred by CambD and Ditto. Compared with EXION, it exploits more sparsity under the quality-preserving SD35 configuration, contributing to its lower energy consumption.
Overall, \T~achieves a {5.5$\times$} average energy saving across the evaluated baselines.

\textbf{Area and Power Analysis.}
Table~\ref{tab:hardwareConfigs} summarizes the hardware characteristics of \T, which occupies 88.69~mm$^2$ and consumes 283.8~W under the DiT-256 workload at 28nm, with a breakdown in Fig.~\ref{fig:speedup_power_area}(c).
The \T~Engine introduces negligible overhead (0.14\% area, 0.5\% power), showing that differential sparsity integrates seamlessly into systolic-array architectures.

In contrast, EXION pays an additional 22\% area overhead for a separate prediction array and triple-buffered PEs, whereas \T~attaches only a lightweight engine without altering the original PE-array dataflow.
The area of \T~is comparable to the TPU, and slightly larger than CambD and Ditto, whose low-precision compute logic trades generation quality for area.
Overall, \T~gains its performance and quality advantages at marginal hardware cost, offering a more favorable performance-cost tradeoff than existing accelerators.

\begin{figure}[tb]
    \centering
    \includegraphics[width=\linewidth]{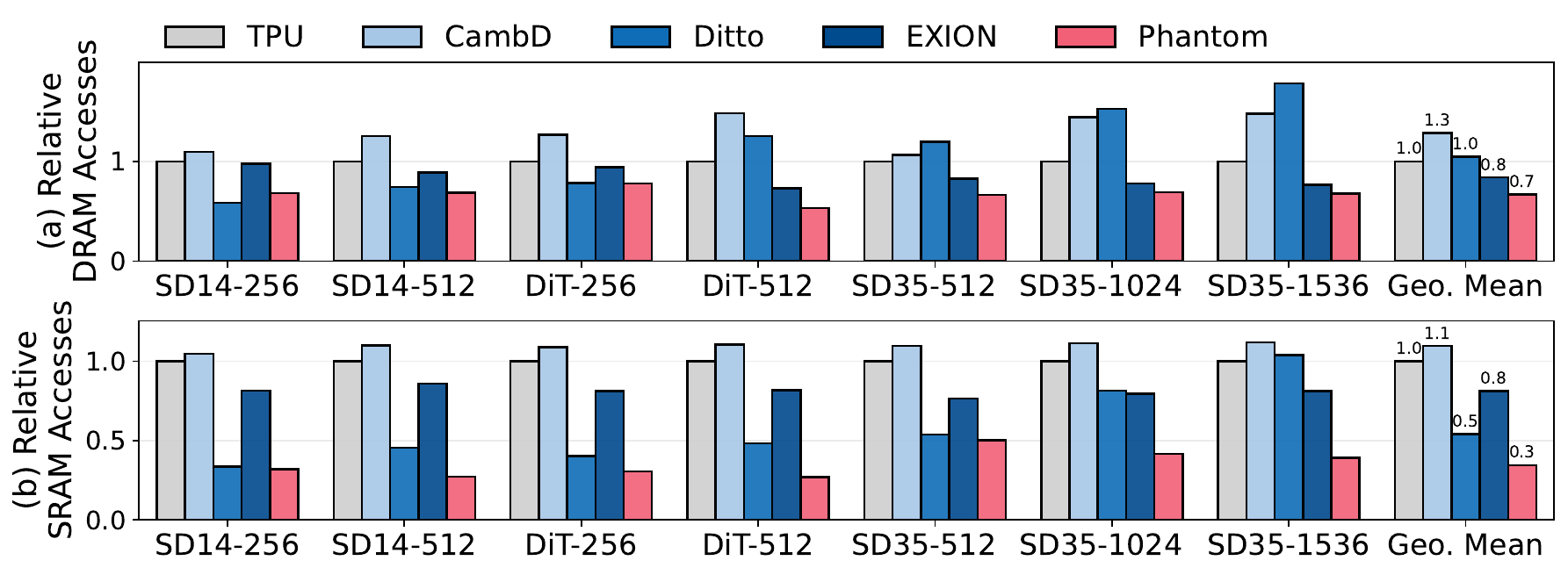}
    \caption{{Relative DRAM/SRAM memory accesses comparison, with TPU as the baseline.}}
    \vspace{-3mm}
    \label{fig:OverallMemCompare}
\end{figure}

\subsection{Memory Analysis}
\label{sec:eval}
\textbf{{Overall DRAM/SRAM.}} Fig.~\ref{fig:OverallMemCompare} compares the DRAM and SRAM accesses of all designs across workloads, measured in total transferred bytes and normalized to the TPU.
On average, \T~reduces DRAM and SRAM accesses by 1.6$\times$ and 2.5$\times$ over all baselines.

Compared with DQ-based designs, \T~reduces DRAM and SRAM accesses by 1.7$\times$ and 2.4$\times$, respectively. The gap has two sources. First, DQ must restore full-precision values around every non-linear function, so each layer boundary incurs an extra round of precision conversion and data movement. Second, its differential attention takes a second pass over the attention data, and this cost grows with model scale. \T~avoids both: it keeps activations at a single precision throughout and preserves the original dense attention dataflow.

Compared with EXION (1.25$\times$ DRAM and 2.35$\times$ SRAM savings), although its FFN cache reduces both computation and memory access, its sparsity prediction requires an extra pass over attention inputs, and its triple-buffered PEs substantially inflate SRAM traffic.
In contrast, \T~optimizes all operations uniformly through the temporal cache without such auxiliary structures.

\begin{figure}[tb]
    \centering
    \subfloat[(a) \textit{Stride} exploration.]{
        \includegraphics[width=0.47\linewidth]{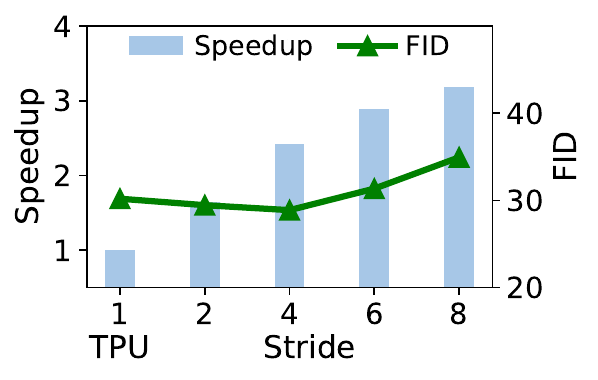}
    }%
    \subfloat[(b) Top\textit{K} exploration.]{
        \includegraphics[width=0.47\linewidth]{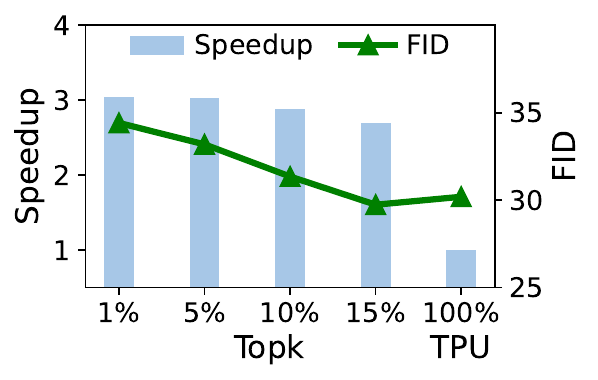}
    }%
    \caption{Design space exploration for the differential-sparsity configurations based on SD35. Speedup is normalized to TPU.}
    \label{fig:Stride_and_Topk}
    \vspace{-3mm}
\end{figure}

\textbf{Temporal Cache Overhead.} We further quantify the cost of the temporal cache itself. \T~caches only the activations needed for differential detection and reuse, so its memory overhead scales as \(O(N)\) with the token length \(N\), rather than the \(O(N^2)\) attention-score dependency of DQ-style~\cite{kim2025ditto} and score-based methods~\cite{zou2024accelerating}. 
Under our hardware configuration, a 4\,MB partition of the global buffer is sufficient for the \T~Engine to perform tile-level detection and reuse with double buffering, so cache accesses fully overlap with PE-array computation. The bulk of the cache resides in DRAM and is fetched on demand, adding only 14.7\% to the total DRAM footprint on average. In terms of traffic, temporal-cache transfers account for 0.25\% of total SRAM accesses and 15.1\% of total DRAM accesses.

\begin{figure}[tb]
    \centering
    \includegraphics[width=0.75\linewidth]{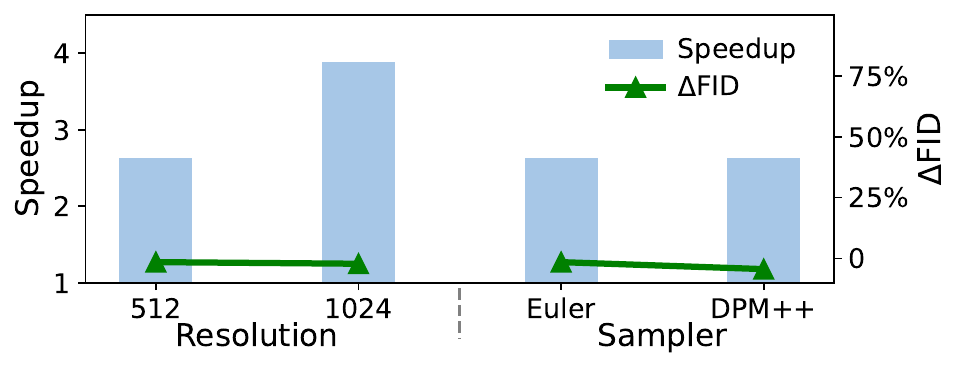}
    \caption{Sensitivity of SD35 to resolution and sampler. Speedup is normalized to the TPU, while $\Delta$FID is measured against the original SD35 model under the corresponding setting.}
    \label{fig:sensitivity}
\end{figure}

\subsection{Ablation Study}

\textbf{Tick-Tock Configuration Exploration.}
We perform a design space exploration of the tick-tock differential sparsity over its two hyperparameters, \textit{stride} and Top\textit{K}, aiming to maximize speedup while preserving baseline-level FID.
Fig.~\ref{fig:Stride_and_Topk} shows their impact on the SD35-512 workload, with the TPU as the baseline: \textit{stride}=1 and Top\textit{K}=100\% both reduce to full computation, i.e., TPU-level performance and quality.
We first sweep the stride (Fig.~\ref{fig:Stride_and_Topk}(a)) with Top\textit{K} fixed at 10\%, similar to the outlier ratios in prior works~\cite{kong2024cambricon, kim2025ditto}.
Small strides operate in a near-lossless regime, where FID fluctuates slightly around the baseline, a behavior also observed in lightly pruned or quantized models. \textit{Stride}=6 marks the boundary of this regime, delivering good speedup with a 5\% FID deviation.

We therefore fix \textit{stride}=6 and sweep Top\textit{K} (Fig.~\ref{fig:Stride_and_Topk}(b)): enlarging Top\textit{K} recovers the FID degradation at the cost of speedup, and Top\textit{K}=15\% restores baseline-level quality with the best remaining speedup.
The trends hold across models, and the configuration derived at 512$\times$512 preserves quality at other resolutions, so we apply it uniformly.
Based on this exploration, we set [\textit{stride}, Top\textit{K}] of [6, 15\%] for SD35, [9, 10\%] for SD14, and [6, 5\%] for DiT. 
With mask accumulation, the average retained token fraction increases to 18.8\% on SD35 and 11.4\% on DiT.

\textbf{Sensitivity Analysis.}
Fig.~\ref{fig:sensitivity} evaluates the selected configuration (\textit{stride}=6, Top\textit{K}=15\%) across resolutions and samplers, reporting speedup over the TPU and $\Delta$FID relative to the corresponding original SD35 model.
\T~maintains near-baseline generation quality in all settings, with $\Delta$FID close to zero.
At a fixed resolution and step count, Euler and DPM++ achieve nearly identical speedups, consistent with their shared transformer-block structure and tensor dimensions.
Increasing the resolution from $512$ to $1024$ improves speedup from 2.64$\times$ to 3.88$\times$, consistent with the greater amount of computation avoided by selective recomputation at higher resolutions.

\begin{table}[t]
\centering
\caption{Speedup and energy saving of temporal-reuse policies on GPU and \T~hardware.}
\label{tab:efficiency}
\setlength{\tabcolsep}{3pt}
\renewcommand{\arraystretch}{1}
\begin{tabular*}{\columnwidth}{@{\extracolsep{\fill}}llrrrr@{}}
\toprule
& & \multicolumn{2}{c}{Speedup} & \multicolumn{2}{c}{Energy Saving} \\
\cmidrule(lr){3-4} \cmidrule(l){5-6}
Hardware & Policy & 256 & 512 & 256 & 512 \\
\midrule
\multirow{4}{*}{GPU}
 & FlashAttention~\cite{dao2024flashattention} & 1.00 & 1.00 & 1.00 & 1.00 \\
 & ToCa~\cite{zou2024accelerating}           & 0.98 & 1.31 & 1.14 & 1.88 \\
 & TaylorSeer~\cite{liu2025taylorseers}     & 1.19 & 1.62 & 1.50 & 2.45 \\
 & \T             & 1.06 & 1.27 & 1.28 & 1.88 \\
\midrule
\multirow{3}{*}{\T}
 & ToCa~\cite{zou2024accelerating}           & 5.78 & 3.15 & 10.58 & 9.05 \\
 & TaylorSeer~\cite{liu2025taylorseers}     & 6.79 & 3.36 & 12.72 & 10.38 \\
 & \textbf{\T}    & \textbf{7.79} & \textbf{5.60} & \textbf{17.84} & \textbf{18.19} \\
\bottomrule
\end{tabular*}
\end{table}

{\textbf{Hardware and Temporal-Reuse Comparison.}
Table~\ref{tab:efficiency} compares the reuse policy of \T~with two representative temporal-reuse methods on the same GPU and the same \T~accelerator, using GPU-FlashAttention~\cite{dao2024flashattention} as the baseline.
ToCa~\cite{zou2024accelerating} selects tokens for recomputation using attention-derived scores, as discussed earlier, whereas TaylorSeer~\cite{liu2025taylorseers} avoids attention scores and instead forecasts features from cached ones via Taylor-series expansion across timesteps.
All three methods use their default configurations, with \T~adopting the same configuration as in Table~\ref{tab:network_performance}, and achieve near-baseline FID on DiT.
On GPU, \T~provides a modest speedup and outperforms ToCa on DiT-256 (1.06$\times$ vs.\ 0.98$\times$), while TaylorSeer achieves the highest speedup. 
On the \T~accelerator, however, the ranking reverses.
TaylorSeer's forecasting requires additional reads and updates of cached features and their finite differences, adding memory traffic, whereas the \T~Engine pipelines detection and overlaps reuse with computation.
As a result, our policy achieves the highest speedups (7.79$\times$/5.60$\times$) and energy savings (17.84$\times$/18.19$\times$) on DiT-256/512, running 1.35$\times$/1.78$\times$ faster than ToCa and 1.15$\times$/1.67$\times$ faster than TaylorSeer on the same hardware.
These results show that the benefit of \T~stems from co-designing both algorithm and architecture.}

\begin{figure}[tb]
    \centering
    \includegraphics[width=0.9\linewidth]{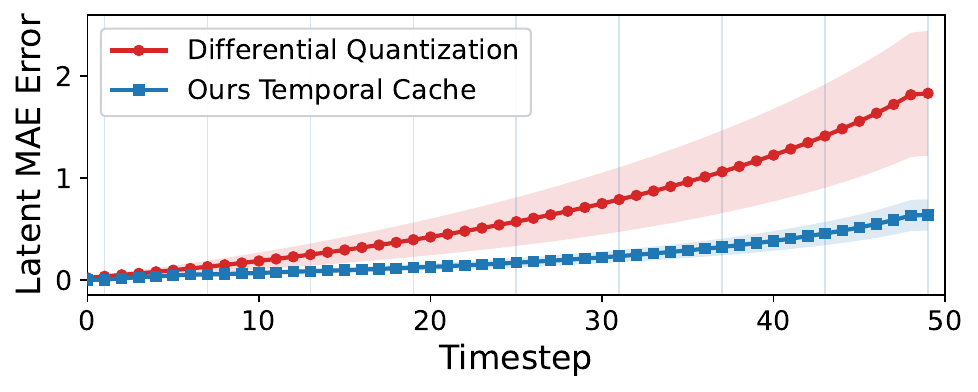}
    \caption{Per-step latent MAE of Differential Quantization (red) and Temporal Cache (blue) on SD35. Shaded regions indicate the variance across prompts from MSCOCO.}
    \label{fig:Error_compare}
    \vspace{-3mm}
\end{figure}

\textbf{Approximation Error: Cache vs. Quantization.}
To quantify and analyze per-step approximation error of \T, we measure the mean absolute error (MAE) of the intermediate latent after each timestep against its FP16 reference, comparing the proposed temporal cache with differential quantization (CambD, INT4) over 50 timesteps on SD35 (averaged over 20 MSCOCO prompts). As shown in Fig.~\ref{fig:Error_compare}, differential quantization injects error at every step, causing the latent MAE to diverge super-linearly. In contrast, the temporal cache introduces only minor error during differential steps, and the tick-tock strategy with periodic fully computation progressively recovers the drift, yielding less error growth. At the final step, the temporal cache achieves a 2.87$\times$ reduction in MAE error, ensuring stable generation quality throughout the diffusion process.

\section{{Conclusion}}

In this work, we present \textbf{\T}, a novel architectural paradigm for accelerating diffusion inference. By reinterpreting temporal similarity as differential sparsity, \T~selectively recomputes sparse outlier activations while reusing the rest from the temporal cache, effectively reducing both computation and memory access.
With the tick-tock differential sparsity, streamed \T~Engine, and condensed pipeline, \T~achieves an average \textbf{{4.0}$\times$} speedup and \textbf{{5.5}$\times$} energy saving over GPUs and prior accelerators while maintaining generation quality, demonstrating its effectiveness for modern diffusion workloads.

\bibliographystyle{IEEEtran}
\bibliography{sample-base}

\end{document}